\documentclass[sii,authoryear]{ipart}
\usepackage{adjustbox,float,amsmath,multirow}
\usepackage{xcolor}

\startlocaldefs
\endlocaldefs

\pubyear{2021}
\volume{0}
\issue{0}
\firstpage{1}
\lastpage{1}

\begin{document}

\begin{frontmatter}

\title{Sparse Logistic Regression on Functional Data\protect\thanksref{}}
\thankstext{T1}{This research was supported in part by the U.S. National Science Foundation grants DMS-1620945 and DMS-1916174.}

\begin{aug}
\author{\inits{Y.}\fnms{Yunnan} \snm{Xu}\ead[label=e1]{yunnan91@vt.edu}},
\address{Novartis International AG,\\1 Health Plaza,\\ East Hanover, NJ 07936\\USA \\\\\printead{e1}}
\author{\inits{P.}\fnms{Pang} \snm{Du}\thanksref{t2}\ead[label=e2]{pangdu@vt.edu}},
\thankstext{t2}{Corresponding author. ORCID: 0000-0003-1365-4831}
\address{Department of Statistics, \\Virginia Tech,\\ 250 Drillfield Drive, \\Blacksburg, VA 24061\\USA \\\printead{e2}}
\author{\inits{J.}\fnms{John} \snm{Robertson}\ead[label=e3]{drbob@vt.edu}}
\address{Department of Biomedical Engineering and Mechanics, \\Virginia Tech,\\ 325 Stanger St., \\Blacksburg, VA 24061\\USA \\\\\printead{e3}}
\and
\author{\inits{R.}\fnms{Ryan} \snm{Senger}\ead[label=e4]{senger@vt.edu}}
\address{Department of Biological Systems Engineering, \\Virginia Tech, \\Blacksburg, VA 24061\\USA \\\\\printead{e4}}
\end{aug}

\begin{abstract} 
Motivated by a hemodialysis monitoring study, we propose a logistic model with a functional predictor, called the Sparse Functional Logistic Regression (SFLR), where the corresponding coefficient function is {\it locally sparse}, that is, it is completely zero on some subregions of its domain. The coefficient function, together with the intercept parameter, are estimated through a doubly-penalized likelihood approach with a B-splines expansion. One penalty is for controlling the roughness of the coefficient function estimate and the other penalty, in the form of the $L_1$ norm, enforces the local sparsity. A Newton-Raphson procedure is designed for the optimization of the penalized likelihood. Our simulations show that SFLR is capable of generating a smooth and reasonably good estimate of the coefficient function on the non-null region(s) while recognizing the null region(s). Application of the method to the Raman spectral data generated from the heomdialysis study pinpoint the wavenumber regions for identifying key chemicals contributing to the dialysis progress.

\end{abstract}




\end{frontmatter}

	\section{Introduction}
\label{se:intr}

In the past decades, functional regression models with functional predictors have attracted a lot of attention from researchers ever since the arrival of the seminal monograph \citet{rbook1}. Among them, functional regression models with a continuous response have been studied the most. Some well-known examples are \citet{yao:05a}, \citet{cai2006}, \citet{hall2007}, \citet{crambes2009}, and \citet{yuan2010}. Such models were later extended to generalized functional linear models (GFLMs) where the response can be discrete such as binary or counts. For example, an early investigation was presented in \citet{james2002} where both continuous and discrete responses were entertained. \citet{muller2005} generalized the functional principal component analysis (FPCA) approach in \citet{yao:05a} to GFLMs. Their approach was further extended to the multi-level functional data scenario by \citet{crainiceanu2009}, and GFLMs with semiparametric single-index interactions in \citet{li2010}. \citet{hall2007} studied the convergence rates for the standard FPCA approach. However, the FPCA approach has a well-known drawback that the functional principal components may not be an efficient basis for representing the coefficient function \citep{yuan2010} and can often produce functional estimate with artificial bumps \citep{rbook}. Therefore, a roughness penalty approach was adopted in \citet{gold11} and \citet{du2014}. In particular, \citet{gold11} approximated the functional predictor by a linear combination of the leading eigenfunctions of the smoothed covariance function and estimated the coefficient function through penalized spline regression. In \citet{du2014}, the coefficient estimator is the exact finite-dimensional optimizer of a penalized likelihood and exhibits the optimal convergence rate for the prediction error. Their approach was extended by \citet{wang2017} to generalized scalar-on-image regression models with a total variation penalty enforced on the regression coefficient function estimator. %
Besides basis expansion and roughness penalty, wavelet representation was also considered for GFLMs. For example, \citet{mous17} extended the work of \citet{zhao12} to a multinomial response and used wavelet representations for predictor functions and the coefficient function. \citet{kayano2016} considered a sparse functional logistic model where an elastic net penalty is used to select signal functional covariates among a large number of available ones.
A common assumption in these existing GFLMs is that the smooth coefficient functions are nonzero on the entire domain (except for possibly a few zero-crossing points). However, this restriction may not be appropriate in some applications where {\it local sparsity} of the regression coefficient function, that is, the function is zero on a subregion or several separate subregions of the domain, has practical meaning and is thus desired. 

\begin{figure}[]
    \centering
    \includegraphics[scale=0.85]{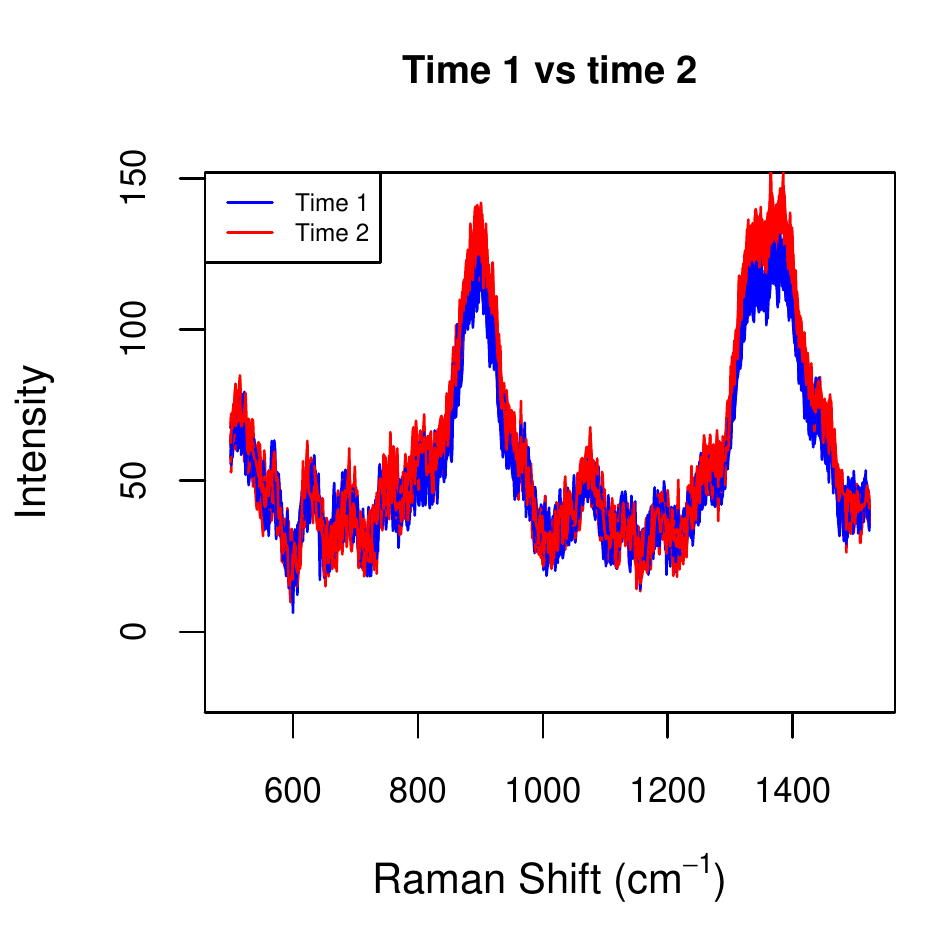}
    \includegraphics[scale=0.85]{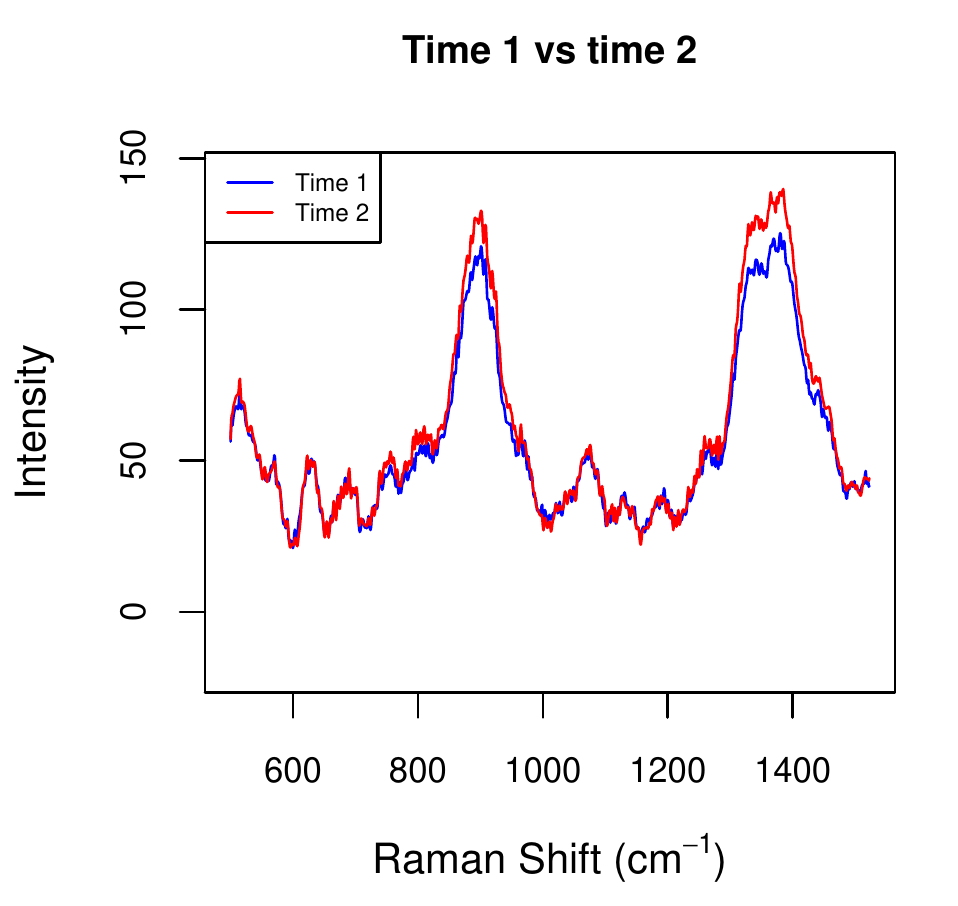}
    \caption{Raman spectra for waste dialysate samples at two time points of a hemodialysis session. Left: Individual spectra. Right: Mean spectra.}
    \label{fig:ch4real}
\end{figure}

Our motivating example comes from a hemodialysis monitoring study. Hemodialysis is a major treatment option for patients with end stage renal diseases. In a hemodialysis treatment session, the dialyzer connected to the patient pumps out the patient's blood and directs it into a chamber containing clean dialysate, a fluid responsible for removing the wastes from the blood. The cleansed blood is directed back to the patient's body while the waste dialysate is discarded through a drainage system. In our experiment, samples of waste dialysate were collected at regularly spaced time points during a standard 4-hour hemodiaysis treatment session. Each sample was divided into 10 portions and each portion was scanned by a Raman spectroscope to produce a Raman spectrum. As an example, Figure~\ref{fig:ch4real} shows two groups of Raman spectra and their mean spectra from waste dialysate samples collected at two different times of a hemodialysis session. Since each spectrum carries critical information about the chemical composition of the corresponding waste dialysate sample, the monitoring procedure naturally demands a comparison of spectra generated at different time points. On one hand, a two-sample test like the one in \citet{Horvath2013} can be used to determine whether the mean spectra at two time points are different or not. On the other hand, it is also important to find out at which regions of the wavenumber domain the mean spectra are different, since the identified range(s) of wavenumbers can reveal which chemicals in the waste dialysate samples cause the difference. These differences in chemical composition can, in turn, be used to determine how efficiently important targeted waste molecules, like urea, are removed. This can be cast as a sparse functional logistic regression problem such that the nonzero region(s) of the regression coefficient function corresponds to the sub-domain contributing to the mean function difference while the zero region(s) corresponds to the sub-domain where the two group mean functions are similar.

Locally sparse functional regression models have been studied when the response is a continuous variable. For example, \citet{james09} proposed a method called  "Functional Linear Regression That’s Interpretable" (FLiRTI). They divided the domain into a large number of sub-intervals such that the problem becomes identify sub-intervals where the coefficient function is zero. Then they expand the coefficient function into a linear combination of locally sparse basis functions and thus reduce the task to a high dimensional variable selection problem for which they adopted the Dantzig selection procedure. Their approach can also incorporate zero regions identifications for the derivatives of the coefficient function. However, the FLiRTI has some drawbacks \citep{zhou13, cao17}. On one hand, it cannot guarantee consecutive zero sub-intervals due to its discretization of the problem and may result in a coefficient function estimate that is hard to interpret. On the other hand, the choice of the number of sub-intervals can be tricky. Precise identification of zero regions would require a large number of sub-intervals but too many sub-intervals can lead to over-parameterization and unstable estimation. To address these issues, \citet{zhou13} proposed a two-stage locally sparse estimator of the coefficient function. They used the Dantzig selector to obtain initial locations of the null sub-regions in the first stage and then refined the location estimates by a group SCAD approach. This method overcomes the underestimation of the non-zero coefficients from the initial Dantzig estimator. Yet the requirement of selecting several tuning parameters at each stage increases estimation variability and computational complexity. More recently, \citet{cao17} proposed a smooth and locally sparse (SLoS) estimator based on a functional extension of the SCAD penalty which regularizes the $L_1$-norm of the estimated coefficient function. B-spline basis functions are employed to facilitate computation due to their compact support property. The SLoS method locates the null subregions and smoothly estimates the non-zero values of the coefficient function without over-shrinkage at the same time. Therefore, apart from reduced variability, the SLoS method also has better interpretability. Despite the above developments, as far as we know, there hasn't been any work on extending locally sparse estimation to the GFLM setting.

In this work we consider the problem of modeling a binary outcome against a functional predictor where the regression coefficient function is locally sparse. Inspired by the SLoS, we introduce a new method called sparse functional logistic regression (SFLR) which applies an $L_1$-norm penalty on the coefficient function to achieve local sparsity as well as a roughness penalty to enforce a certain level of smoothness. We use B-splines to model the coefficient function and a Newton-Raphson procedure to optimize the doubly penalized likelihood for obtaining the estimate. The proposed method produces a smooth estimate of the coefficient function that recognizes all the null regions. It has the following distinct features: (1) it is the first GFLM that incorporates local sparsity of the coefficient function into the estimation; (2) the null regions it identifies has important practical meaning and represent where the two groups are similar, while the non-null regions can provide key information for differentiating the two groups; (3) its computation is convenient with the Newton-Raphson procedure even with the complexity of double penalties. We test the SFLR on simulated data under different settings in terms of the misclassification rates, sensitivity, specificity, and prediction errors. Its application to the hemodialysis monitoring study yield critical regions for researchers to identify key chemicals in the waste dialysate samples. 

The rest of the paper is laid out as follows. Section \ref{se:meth} explains SFLR method, and covers the computational details. Section \ref{se:sim} and \ref{se:real} shows the performance of SLR on simulated data and real data. Section \ref{se:con} summarizes the proposed method.

\section{Method}
\label{se:meth}

\subsection{Model and Objective Function}
\label{sse:mod}

Suppose that we have independent observations $(y_i, x_i(t)), i=1,\ldots,N$ where $y_i\in \{0,1\}$ is the binary response and $x_i(t)$ is a square-integrable function defined on a compact interval $\mathcal{T}$, which we assume, without loss of generality, to be $[0,T]$ for some $T>0$. Assume that $y_i \sim \text{Bernoulli}(p_i)$ with $p_i=\text{Prob}(y_i=1)$ following the functional logistic regression model
\begin{equation}
\label{eq:flmodel}
\log\left(\frac{p_i}{1-p_i}\right)=\alpha + \int^T_0 \beta(t) x_i(t) dt,
\end{equation} 
where $\alpha$ is the intercept and $\beta(t)$ is a smooth coefficient function. In particular, we are interested in the case that $\beta(t)$ is {\it locally sparse}, that is, there exists an unknown subregion $\mathcal{Z}$ of $[0,T]$ such that $\beta(t)=0$ for all $t\in \mathcal{Z}$. Intuitively, $\mathcal{Z}$ represents the region where the predictor process $x(t)$ carries no information about the binary response $y$. Therefore, the identification of $\mathcal{Z}$ is of importance parallel to the estimation of $\beta(\cdot)$.

The log-likelihood function for model \eqref{eq:flmodel} is 
\begin{equation}
\label{eq:logl}
\begin{aligned}
   l(\beta)&= \sum_{i=1}^N \Bigl(y_i\{\alpha + \int^T_0 \beta(t) x_i(t) dt\} \\
& -\log[1+\exp\{\alpha + \int^T_0 \beta(t) x_i(t) dt\}]\Bigr). 
\end{aligned}
\end{equation}  

To estimate the smooth and locally sparse coefficient function $\beta(t)$, we need an $L_1$-norm penalty $\int^T_0 |\beta(t)|dt$  for local sparsity control and a roughness penalty on $\beta(t)$ for smoothness guarantee. Therefore, our final objective function is the penalized likelihood
\begin{equation}
    J(\beta)=-l(\beta)+\gamma\int^T_0 \{\beta^{(m)}(t)\}^2dt+\lambda\int^T_0 |\beta(t)|dt,
\label{eq:obf}
\end{equation}
where $\gamma$ and $\lambda$ are positive tuning parameters weighing respectively the smoothness and local sparsity of $\beta$.
The order $m$ of derivative in \eqref{eq:obf} also specifies the order of splines for modeling $\beta$. For example, $m=2$ would correspond to cubic splines. Our estimator of $\beta(t)$ is defined by 
\begin{equation}
    \begin{aligned}
     \hat{\beta}(t) &= \arg\min_{\beta} \Big\{-\sum_{i=1}^N \Bigl(y_i\{\alpha + \int^T_0 \beta(t) x_i(t) dt\}\\
    &-\log[1+\exp\{\alpha + \int^T_0 \beta(t) x_i(t) dt\}]\Bigr) \\
    & + \gamma\int^T_0 \{\beta^{(m)}(t)\}^2dt+\lambda\int^T_0 |\beta(t)|dt\Big\}.
\end{aligned}
\label{eq:best}
\end{equation}
Note that we do not introduce a notation for the estimator of $\alpha$ here only for the simplicity of presentation. The intercept $\alpha$ can be naturally incorporated into the estimation of $\beta$ once $\beta$ is expressed as a linear combination of spline basis functions. 

\subsection{Computation}
\label{sse:comp}

We shall optimize the objective function in \eqref{eq:best} through the Newton-Raphson procedure. This involves first rewriting the objective function in a matrix-vector format after approximating the coefficient by a B-spline basis expansion, deriving a local quadratic approximation to the $L_1$-norm sparsity penalty, and deriving the updating equation for the Newton-Raphson procedure. 

We represent the coefficient function $ \beta(t)$ by B-splines defined on $[0, T]$ as 
\begin{equation}
    \beta(t) \approx \sum_{l=1}^L b_l e_l(t) = \boldsymbol{e}^{\rm T}(t)\boldsymbol{b},
     \label{eq:bsp}
\end{equation}
where $\boldsymbol{e}(t)=(e_1(t), \ldots, e_L(t))^{\rm T}$ is a set of $L$ order-$(d+1)$ B-spline basis functions, and $\boldsymbol{b}=(b_1, \ldots, b_L)^{\rm T}$ is the coefficient vector.
Suppose $e_l(t)$'s are defined by $(M+1)$ equally spaced knots, $0=t_0 <t_1 < \ldots < t_M = T$. Two consecutive knots make a sub-interval on which $e_l(t)$ is a polynomial of degree $d$. The compact support property guarantees each B-spline basis function is nonzero only on at most $(d+1)$ sub-intervals in succession which form a small sub-region when $M$ is large. Then 
the log likelihood \eqref{eq:logl} is rewritten as
\begin{equation}
\begin{aligned}
   l(\mathbf{b}) &= \sum_{i=1}^N \Bigl(y_i\{\alpha + \int^T_0 x_i(t)\boldsymbol{e}^{\rm T}(t) dt\, \boldsymbol{b})\}\\
   &-\log\Big[1+\exp\big\{\alpha + \int^T_0 x_i(t)\boldsymbol{e}^{\rm T}(t) dt\, \boldsymbol{b}\big\}\Big] \Bigr)   \\
    &= \sum_{i=1}^N \left[y_i(\alpha + \boldsymbol{U}^{\rm T}_i \boldsymbol{b}) -\log\{1+\exp(\alpha + \boldsymbol{U}^{\rm T}_i \boldsymbol{b})\}\right],
\end{aligned}
\label{eq:loglb}
\end{equation}
 where $\boldsymbol{U}_i=\int^T_0 x_i(t)\boldsymbol{e}^{\rm T}(t) dt$, and $\boldsymbol{U}=(\mathbf{U}_1,\ldots,\mathbf{U}_N)^{\rm T}$ is an $N$ by $L$ matrix.
  The roughness penalty in \eqref{eq:best} can be rewritten as
  \begin{equation}
      \label{eq:rpquad}
      \gamma\int^T_0 \{\beta^{(m)}(t)\}^2dt=\gamma \boldsymbol{b}^{\rm T}\boldsymbol{V}\boldsymbol{b},
  \end{equation}  
 where $\boldsymbol{V}$ is an $L$ by $L$ matrix with $v_{ij}=\int_0^T (\frac{d^m e_i(t)}{dt^m}\frac{d^m e_j(t)}{dt^m})dt $ for $1\leq i,j\leq L$.

Next, we need to derive a local quadratic approximation to the sparsity penalty.
Let $p_\lambda (|\beta(t)|)= \lambda |\beta(t)|$.
By Theorem 1 in \citet{cao17}, the sparsity penalty in \eqref{eq:best} can be expressed as
\begin{equation}
    \int^T_0 p_\lambda (|\beta(t)|)dt = \frac{T}{M}\sum^M_{j=1} p_\lambda \left(\frac{\|\beta_{[j]}\|_2}{\sqrt{T/M}}\right)
\label{eq:lasso}
\end{equation}
where $\|\beta_{[j]}\|_2=\sqrt{ \int^{t_j}_{t_{j-1}}\beta^2(t)dt}$.
The Taylor expansion at the current estimate $\Tilde{\beta}$ gives
\begin{equation}
    \begin{aligned}
    \sum^M_{j=1} p_\lambda \left(\frac{\|\beta_{[j]}\|_2}{\sqrt{T/M}}\right)& \approx \frac{1}{2}\sum^M_{j=1} \frac{p_\lambda' \left(\frac{\|\Tilde{\beta}_{[j]}\|_2}{\sqrt{T/M}}\right)}{\frac{\|\Tilde{\beta}_{[j]}\|_2}{\sqrt{T/M}}} \frac{\|\beta_{[j]}\|_2^2}{T/M} + G(\Tilde{\beta})
    \\
    & = \frac{\lambda}{2\sqrt{T/M}}\sum^M_{j=1} \frac{1}{\|\Tilde{\beta}_{[j]}\|_2} \|\beta_{[j]}\|_2^2 + G(\Tilde{\beta}),
\end{aligned}
    \label{eq:taylorp}
\end{equation}
where $G(\Tilde{\beta})$ does not depend on $\beta(t)$. Note that 
\begin{equation}
    \|\beta_{[j]}\|_2^2 =\int^{t_j}_{t_{j-1}}\beta^2(t)dt 
    = \boldsymbol{b}^{\rm T}\int^{t_j}_{t_{j-1}}\boldsymbol{e}(t)\boldsymbol{e}^{\rm T}(t)dt \boldsymbol{b} 
    =\boldsymbol{b}^{\rm T}\boldsymbol{W}_j\boldsymbol{b},
\label{eq:bjb}
\end{equation}
 where $\boldsymbol{W}_j=\int^{t_j}_{t_{j-1}}\boldsymbol{e}(t)\boldsymbol{e}^{\rm T}(t)dt$ is an $L$ by $L$ matrix with $w_{uv}=\int^{t_j}_{t_{j-1}} e_u(t) e_v(t) dt $ for $j\leq u,v\leq j+d$ and 0 otherwise. 
Plugging \eqref{eq:bjb} in \eqref{eq:taylorp} gives 
\begin{align*}
    \sum^M_{j=1} p_\lambda \left(\frac{\|\beta_{[j]}\|_2}{\sqrt{T/M}}\right)
    &= \frac{\lambda}{2\sqrt{T/M}} \mathbf{b}^{\rm T} \left(\sum^M_{j=1} \frac{1}{\|\Tilde{\beta}_{[j]}\|_2}\mathbf{W}_j\right)\mathbf{b} + G(\Tilde{\beta})\\
    &=\frac{\lambda}{2\sqrt{T/M}} \mathbf{b}^{\rm T} \Tilde{\mathbf{W}}\mathbf{b} + G(\Tilde{\beta}),
\end{align*}
where $\boldsymbol{\Tilde{W}}=\sum^M_{j=1} \|\Tilde{\beta}_{[j]}\|_2^{-1}\boldsymbol{W}_j$. Therefore, 
\begin{equation}
\begin{aligned}\lambda\int^T_0 |\beta(t)|dt &\approx \frac{\lambda\sqrt{T/M}}{2}\mathbf{b}^{\rm T} \Tilde{\mathbf{W}}\mathbf{b} + \frac{\lambda\sqrt{T/M}}{2}G(\Tilde{\beta})
\end{aligned}
\label{eq:pen}
\end{equation}

Combining \eqref{eq:loglb}, \eqref{eq:rpquad}, and \eqref{eq:pen}, the objective function in \eqref{eq:best}, after dropping the terms not dependent on $\mathbf{b}$, becomes
\begin{equation}
\begin{aligned}
    J(\mathbf{b})&=-\sum_{i=1}^N \left[y_i(\alpha + \mathbf{U}^{\rm T}_i\mathbf{b})-\log\{1+\exp(\alpha + \mathbf{U}^{\rm T}_i\mathbf{b})\}\right]\\
    &+  \mathbf{b}^{\rm T}\mathbf{V}^* \mathbf{b}+ \mathbf{b}^{\rm T} \Tilde{\mathbf{W}}^{*}\mathbf{b},
\end{aligned}
\label{eq:taylor}
\end{equation}
where $\boldsymbol{\Tilde{W}}^{*}=\frac{\lambda\sqrt{T/M}}{2} \boldsymbol{\Tilde{W}}$, and $\boldsymbol{V}^*=\gamma \boldsymbol{V}$.
Derivatives of $J$ are
\begin{equation}
\begin{aligned}
    \frac{\partial J(\mathbf{b})}{ \partial \boldsymbol{b}}
    &= -\sum_{i=1}^N \boldsymbol{U}_i[y_i - P(\boldsymbol{U}_i;\boldsymbol{b},\alpha)] +
    \boldsymbol{V}^* \boldsymbol{b}+ \boldsymbol{\Tilde{W}}^{*} \boldsymbol{b}\\
    &=-\boldsymbol{U}^{\rm T} (\boldsymbol{y} - \boldsymbol{c})  + \boldsymbol{V}^* \boldsymbol{b}+  \boldsymbol{\Tilde{W}}^{*} \boldsymbol{b}
    ,
    \\
    \frac{\partial^2 J(\mathbf{b})}{ \partial \boldsymbol{b} \partial \boldsymbol{b}^{\rm T}}
    &= \sum_{i=1}^N \boldsymbol{U}_i\boldsymbol{U}_i^{\rm T} P(\boldsymbol{U}_i;\boldsymbol{b},\alpha)[1 - P(\boldsymbol{U}_i;\boldsymbol{b},\alpha)] + \boldsymbol{V}^* + \boldsymbol{\Tilde{W}}^{*}\\
    &=\boldsymbol{U}^{\rm T}\boldsymbol{D}\boldsymbol{U}  + \boldsymbol{V}^* + \boldsymbol{\Tilde{W}}^{*},
\end{aligned}
\label{eq:derivative}
\end{equation}
where $P(\boldsymbol{U}_i;\boldsymbol{b},\alpha)= \frac{\exp(\alpha + \boldsymbol{b}^{\rm T}\boldsymbol{U}_i)}{1+\exp(\alpha + \boldsymbol{b}^{\rm T}\boldsymbol{U}_i)}$,  $\boldsymbol{y}=(y_1,\ldots, y_N)^{\rm T}$, $\boldsymbol{c}=(P(\boldsymbol{U}_1;\boldsymbol{b},\alpha), \ldots, P(\boldsymbol{U}_N;\boldsymbol{b},\alpha))^{\rm T}$, and $\boldsymbol{D}=\text{diag}(d_{i,i}), 1\leq i \leq N$, with $ d_{i,i}= P(\boldsymbol{U}_i;\boldsymbol{b},\alpha)(1-P(\boldsymbol{U}_i;\boldsymbol{b},\alpha))$.

Therefore, the Newton-Raphson updating step is 
\begin{equation}
    \begin{aligned}
    \boldsymbol{b}^{(new)}
    &= \boldsymbol{b}^{(old)} - \left(\frac{\partial^2 J(\beta)}{ \partial \boldsymbol{b} \partial \boldsymbol{b}^{\rm T}}\right)^{-1} \frac{\partial J(\beta)}{ \partial \boldsymbol{b}}\\
    &= \boldsymbol{b}^{(old)} + (\boldsymbol{U}^{\rm T}\boldsymbol{D}\boldsymbol{U}  + \boldsymbol{V}^*+ \boldsymbol{\Tilde{W}}^{*})^{-1}\\
    &[\boldsymbol{U}^{\rm T} (\boldsymbol{y} - \boldsymbol{c}) -\boldsymbol{V}^*\boldsymbol{b}^{(old)}  -  \boldsymbol{\Tilde{W}}^{*}\boldsymbol{b}^{(old)}],
\end{aligned}
\label{eq:npupdate}
\end{equation}
where $\boldsymbol{D}$ and $\boldsymbol{c}$ are calculated based on $\boldsymbol{b}^{(old)}$, $\boldsymbol{V}^*$ and $\boldsymbol{W}_j$ are calculated based on the B-spline basis functions before iterations, and $\boldsymbol{\Tilde{W}}^{*}$ is updated in each iteration.

The complete algorithm consists of the following steps.

\begin{itemize}
    \item 
    Step 1: Obtain an initial estimate $\hat{\boldsymbol{b}}^{(0)}$ through the optimization of \eqref{eq:obf} with the sparsity penalty removed. This corresponds to a penalized B-spline estimate of $\beta$.
    \item
    Step 2: During each iteration, update $\hat{\boldsymbol{b}}$ through formula \eqref{eq:npupdate} and then update $\boldsymbol{\Tilde{W}}^{*}$. 
\item
Step 3: Repeat step 2 until convergence. Entries in $\hat{\boldsymbol{b}}$ that are smaller than a threshold $\epsilon$ are set to 0.
\item
Step 4: The output $\hat{\boldsymbol{b}}$ is then used to compute the the estimate of coefficient function by $\hat{\beta}(t)=\boldsymbol{e}^{\rm T}(t)\hat{\boldsymbol{b}}$.
\end{itemize}
To prevent numerical instability due to the probabilities $P(\mathbf{U}_i; \mathbf{b}, \alpha)$ close to 0 or 1, we also set a threshold $\delta$ in Step 2 such that any probabilities falling below $\delta$ are set to $\delta$ and any probabilities going beyond $1-\delta$ are set to $1-\delta$. In this paper, we use the thresholds $\delta=10^{-5}$ and $\epsilon=10^{-4}$.
For the B-splines, we use the cubic splines and around 30 basis functions with equally-spaced knots unless otherwise specified. The tuning parameters $\lambda$ and $\gamma$ can be selected through cross-validation (CV), the Bayesian information criterion (BIC), or the Akaike information criterion (AIC). The parameter $M$ dictates the number of basis functions in the B-splines expansion of the coefficient function. In general, when a roughness penalty is used the choice of $M$ is not crucial so long as it is sufficiently large; see, e.g., Chapter 5 of \citet{rbook}. The simulations in \citet{cao17} further demonstrates that a large $M$ performs better than a small $M$ in identifying the null regions of the coefficient function. For the exact choice of $M$, we follow the guideline in \citet{kg04} and use $M=\max(30, 10n^{2/9})$, where $n$ is the number of discrete sampling points for a functional predictor.

\section{Simulation Studies}
\label{se:sim}
We consider two kinds of predictor functions $X_i(t)$ in our simulation studies. In the first setting they were generated from common functions. In the other setting they were generated from functions resembling Raman spectra.
BIC was used to select tuning parameters unless specified otherwise.

\subsection{Simulation Setting 1: Common Functions as Observed Data}
\label{se:sim1}
\begin{figure}[]
    \centering
    \includegraphics[scale=0.85]{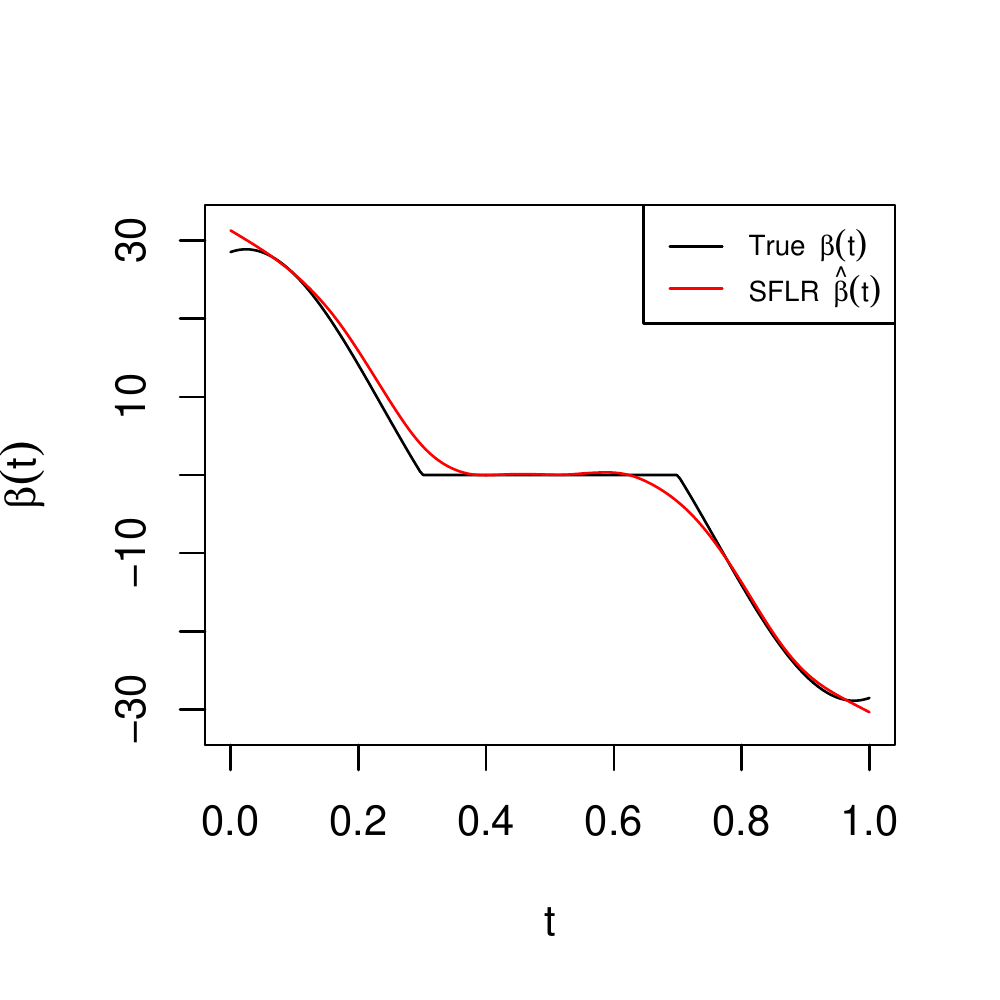}
    \includegraphics[scale=0.85]{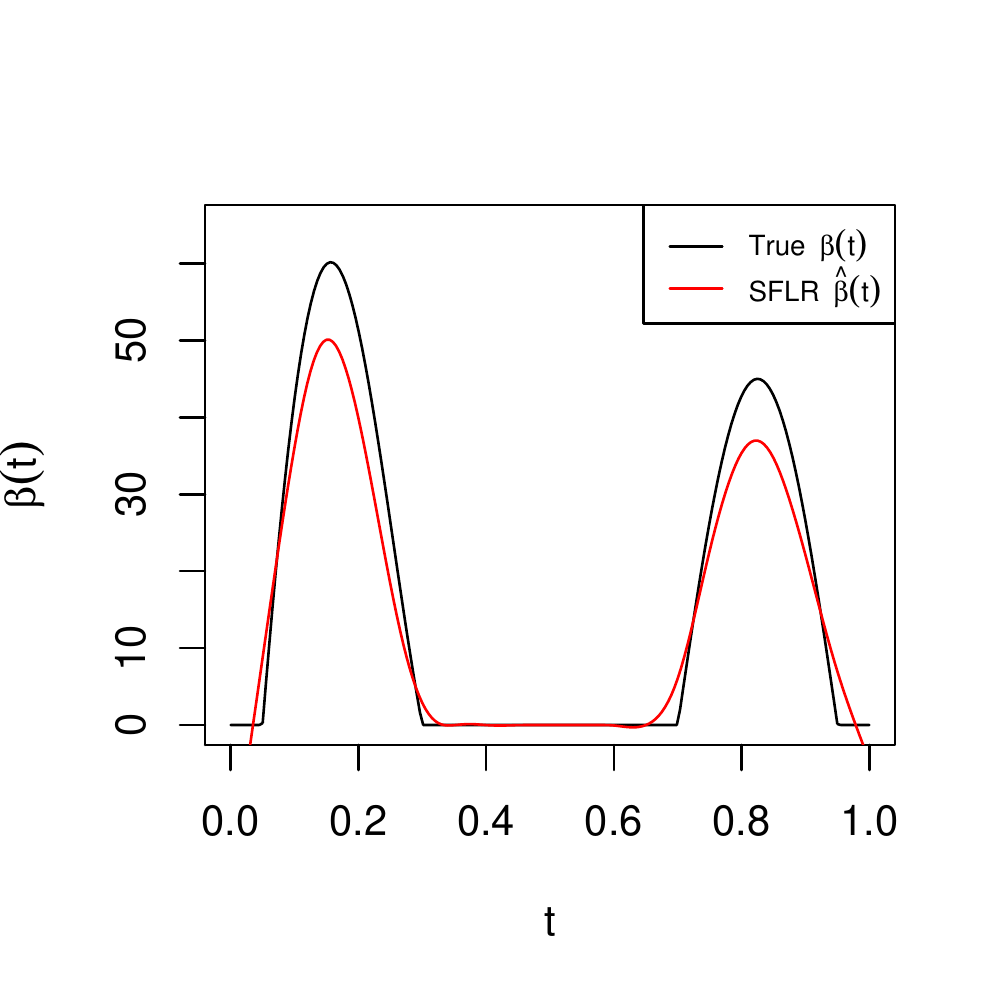}
    \caption{True $\beta(t)$ and SFLR estimates of $\beta(t)$ for simulations with $N=1000$ in Section~\ref{se:sim1}.}
    \label{fig:sflr_beta5}
\end{figure}

 In this section, the proposed SFLR method is tested with different types of coefficient function $\beta(t)$ and sample sizes. We considered two types of $\beta(t)$. The first type, shown in \eqref{eq:beta1}, contains one zero region. The second type, shown in \eqref{eq:beta2}, contains three zero regions. True functions of both types are plotted as solid black lines in Figure \ref{fig:sflr_beta5}.
\begin{equation}
 \beta(t)=\left\{
  \begin{array}{l l}
    15(1-t)\sin(2\pi(t+0.2)),       & \quad \text{if } 0\leq t \leq 0.3,\\
    0, & \quad \text{if } 0.3 < t < 0.7,\\
    15 t\, \sin(2\pi(t-0.2)),  & \quad \text{if } 0.7\leq t \leq 1.
  \end{array}\right.
    \label{eq:beta1}
\end{equation}

\begin{equation}
 \beta(t)=\left\{
  \begin{array}{l l}
      0, & \quad \text{if } 0\leq t < 0.05,\\
    180(t-0.5)\sin(4\pi(t+0.7)),       & \quad \text{if } 0.05\leq t \leq 0.3,\\
    0, & \quad \text{if } 0.3 < t < 0.7,\\
    45t \, \sin(4\pi(t+0.3)),  & \quad \text{if } 0.7\leq t \leq 0.95,\\
    0, & \quad \text{if } 0.95< t \leq 1.
  \end{array}\right.
    \label{eq:beta2}
\end{equation}

 We used model \eqref{eq:flmodel} with $\alpha=0$ to simulate the data. The standard normal distribution was used to generate the coefficient matrix $\boldsymbol{B_x}$ for B-spline basis functions. Then the covariate functions $\mathbf{X}(t)$ were obtained through $\mathbf{X}(t)=\boldsymbol{B_x}\boldsymbol{e}(t)$, where $\boldsymbol{e}(t)$ is a set of 74 order-5 B-spline basis functions with 71 equally spaced knots. The responses were generated from the functional logistic regression model \eqref{eq:flmodel} with $\alpha=0$. Through these steps we generated a training dataset and an independent test dataset. The sample size of the test dataset was kept at 1000 while the training dataset had sample sizes of 50, 150, 450 or 1000.

 Both estimation and prediction performance were assessed. 
 The prediction performance was evaluated on test datasets using the misclassification rate (MCR), sensitivity, specificity, false discovery rate (FDR), and prediction mean squared errors (PMSE). 
 The PMSE was calculated from the predicted probabilities for the test dataset as $PMSE=\frac{1}{N}\sum_i^N (p_i-\hat{p}_i)^2$.
The integrated squared errors (ISE) was used to measure the estimation quality of $\hat{\beta}(t)$. Following \cite{cao17}, we considered two components of the ISE: $ISE_0=\frac{1}{l_0}\int_{\mathcal{Z}} (\hat{\beta}(t)-\beta(t))^2 dt$ defined on the null region $\mathcal{Z}$, and $ISE_1=\frac{1}{l_1}\int_{\mathcal{T}\backslash\mathcal{Z}} (\hat{\beta}(t)-\beta(t))^2 dt $ defined on the non-null region $\mathcal{T}\backslash\mathcal{Z}$, where $l_0$ and $l_1$ are respectively the total lengths of the null and non-null regions. 
When one-null-region $\beta$ was considered, the parameters $\lambda$ and $\gamma$ were respectively selected from the grids $(0.4,0.5,0.6,0.7)*17$ and $(1e-5,1e-6)*15$. When two-null-region $\beta$ was considered, the parameters $\lambda$ and $\gamma$ were respectively selected from the grids
$(0.6,0.7,0.8,0.9,0.95,1)*17$ and $(1e-5,1e-6,1e-7,5e-8)*15$.
For each simulation scenario, we applied the proposed method 
to 100 replications and calculated the medians for each assessment criterion.

We first investigated the choice of the tuning procedure by comparing the 5-fold cross-validation (5-CV), AIC and BIC for the one-null-region $\beta$ with sample sizes $N=450$ and 1000. The results are summarized in Tables \ref{tab:tune1} and \ref{tab:tune2}. Clearly, the AIC and BIC had similar performance. The 5-CV had smaller $ISE_0$ but at the cost of larger $ISE_1$. Also, it took much longer time when the sample size was big. Therefore, we will use the BIC as the tuning procedure from now on.

\begin{table}[ht]
\centering
\begin{adjustbox}{width=\columnwidth,center}
\begin{tabular}{rrrrrrrr}
  \hline
 & MCR & Sensitivity & Specificity & FDR & $ISE_0$ & $ISE_1$ & PSME \\ 
  \hline
BIC & 0.2370 & 0.7654 & 0.7606 & 0.2346 & 0.9043 & 18.8011 & 2.7402 \\ 
  AIC & 0.2370 & 0.7654 & 0.7606 & 0.2346 & 0.9043 & 18.8011 & 2.7402 \\ 
  5-CV & 0.2360 & 0.7626 & 0.7605 & 0.2374 & 0.4804 & 34.4718 & 2.7760 \\ 
   \hline
\end{tabular}
\end{adjustbox}
\caption{Simulation comparison of tuning procedures for $N=450$. Numbers are medians of the assessment criteria.}
\label{tab:tune1}
\end{table}
\begin{table}[ht]
\centering
\begin{adjustbox}{width=\columnwidth,center}
\begin{tabular}{rrrrrrrr}
  \hline
 & MCR & Sensitivity & Specificity & FDR & $ISE_0$ & $ISE_1$ & PSME \\ 
  \hline
BIC & 0.2365 & 0.7657 & 0.7611 & 0.2343 & 0.6766 & 12.3729 & 2.7083 \\ 
  AIC & 0.2365 & 0.7657 & 0.7611 & 0.2343 & 0.6704 & 12.3729 & 2.7083 \\ 
  5-CV & 0.2350 & 0.7659 & 0.7612 & 0.2341 & 0.5107 & 13.9746 & 2.7148 \\ 
   \hline
\end{tabular}
\end{adjustbox}
\caption{Simulation comparison of tuning procedures for $N=1000$. Numbers are medians of the assessment criteria.}
\label{tab:tune2}
\end{table}

Tables \ref{tab:b1ns} and \ref{tab:b5ns} respectively summarize the calculated medians of all the criteria
for $\beta(t)$ from \eqref{eq:beta1} and \eqref{eq:beta2} with sample size at 50, 150, 450 and 1000. In both scenarios, the performance of the proposed method, in terms of all the prediction and estimation criteria, clearly improved as the sample size increased. The prediction performance was satisfactory with both MCR and FDR around 20\% and both sensitivity and specificity around 70-80\%. Overall the prediction performance in the second scenario was slightly better than that in the first scenario. For the estimation performance, the ISE\textsubscript{0} in the first scenario were all close to 0, indicating accurate identification of the null region. The ISE\textsubscript{0} in the second scenario, however, seemed to be higher than expected. This might be caused by the two small null subregions of $\beta(t)$ on the ends of the domain where accurate smoothing to zero can be hard due to less data available there. Plots of two example estimates of $\beta(t)$ in Figure \ref{fig:sflr_beta5} provide further evidence for these conclusions. 

\begin{table}[]
\centering
\begin{adjustbox}{width=\columnwidth,center}
\begin{tabular}{cccccccc}
   \hline
 Sample size & MCR & Sensitivity & Specificity & FDR & $ISE_0$ & $ISE_1$ & PMSE \\ 
  \hline
$N=50$& 0.2805 & 0.7167 & 0.7175 & 0.2833 & 0.2880 &  $180.4400$ &  $3.0490 $\\ 
 \hline
 $N=150$&  $0.2420$  &  $0.7572$ &  $0.7523$ &  $0.2428$ & 0.4255 & $ 57.9419$ &  $2.8513$ \\ 
   \hline
 $N=450$&  
  0.2370 & 0.7654 & 0.7606 & 0.2346 & 0.9043 & 18.8011 & 2.7402\\ 
   \hline
$N=1000$  & 
  $0.2360$ &  $0.7645$ &  $0.7636$ & $0.2355 $ & 0.6008 &  $9.5085$ &  $2.7024$ \\ 
 \hline
\end{tabular}
\end{adjustbox}
\caption{Simulation performance with one-null-region $\beta$ (Section~\ref{se:sim1}). Numbers are medians of the assessment criteria.}
\label{tab:b1ns}
\end{table}

\begin{table}[]
\centering
\begin{adjustbox}{width=\columnwidth,center}
\begin{tabular}{cccccccc}
   \hline 
 Sample size & MCR & Sensitivity & Specificity & FDR & $ISE_0$ & $ISE_1$ & PMSE \\ 
  \hline
n=50&  0.2400 & 0.7636 & 0.7639 & 0.2364 & 186.7040 & 605.1388 & 8.9227 \\ 
  \hline
n=150  & 0.1790 & 0.8220 & 0.8212 & 0.1780 & 103.6906 & 424.2930 & 8.5331 \\ 
  \hline
n=450  & 0.1630 & 0.8381 & 0.8378 & 0.1619 & 43.4276 & 144.1832 & 8.1153 \\ 
\hline
n=1000  & 0.1610 & 0.8394 & 0.8398 & 0.1606 & 19.5676 & 50.1449 & 8.0885 \\ 
 \hline
\end{tabular}
\end{adjustbox}
\caption{Simulation performance with three-null-region $\beta$ (Section~\ref{se:sim1}). Numbers are medians of the assessment criteria.}
\label{tab:b5ns}
\end{table}

To investigate how the method performs with noisier data, we considered a noisier simulation for the case of one-null-region $\beta$. In particular, we added some noise to the predictor functions with a signal-to-noise ratio of 1. The results are summarized in Table~\ref{tab:snr50}. Compared with the results in Table~\ref{tab:b1ns}, we can see that the performance got worse as expected due to the extra noise but still was reasonably good.
\begin{table}[ht]
\centering
\begin{adjustbox}{width=\columnwidth,center}
\begin{tabular}{cccccccc}
   \hline 
 Sample size & MCR & Sensitivity & Specificity & FDR & $ISE_0$ & $ISE_1$ & PMSE \\ 
  \hline
$N=50$ & 0.3970 & 0.6055 & 0.6020  & 0.3945 & 0.0000 & 431.0153 & 3.6060 \\ 
  \hline
$N=150$ & 0.2640  & 0.7363 & 0.7377 & 0.2637 & 0.0824 & 142.1826 & 3.0460 \\ 
  \hline
$N=450$ & 0.2400 & 0.7614 & 0.7587 & 0.2386 & 1.0062 & 86.8439 & 2.8879 \\ 
  \hline
$N=1000$ & 0.2380 & 0.7630 & 0.7607 & 0.2370 & 0.5804 & 75.6508 & 2.8793 \\ 
   \hline
\end{tabular}
\end{adjustbox}
\caption{Simulation performance with noisier data for one-null-region $\beta$ (Section~\ref{se:sim1}). Numbers are medians of the assessment criteria.}
    \label{tab:snr50}
\end{table}

\subsection{Simulation Setting 2: Spectral Data as Observed Data}
\label{se:sim2}

\begin{figure}[H]
    \centering
    \includegraphics[scale=0.75]{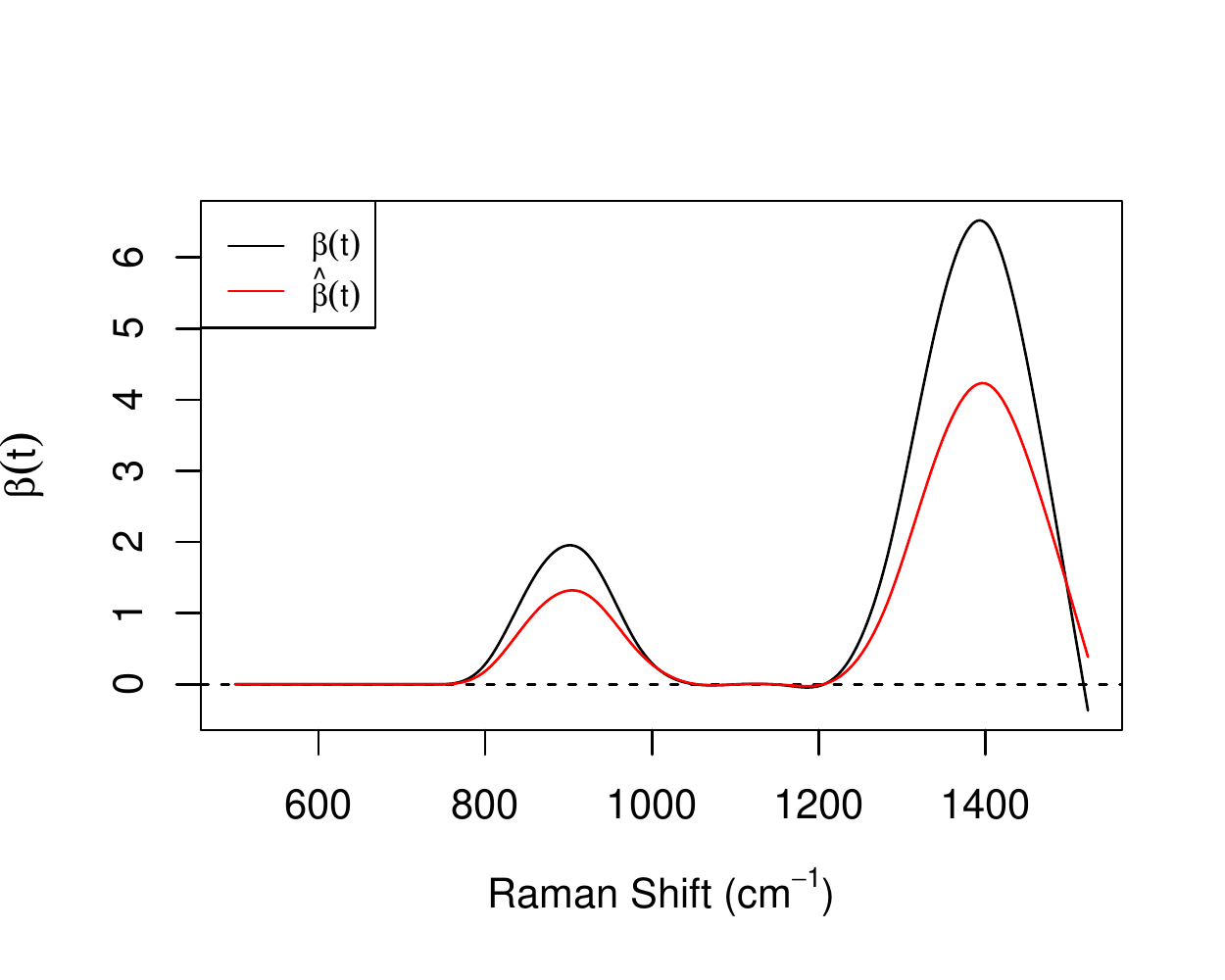}
    \caption{True $\beta(t)$ and estimated $\hat{\beta}(t)$ for simulations in Section~\ref{se:sim2}. }
    \label{fig:sflr_sim2b}
\end{figure}

We also did a single replicate simulation with predictor functions selected to mimic the Raman spectra in our application. We generated covariate functions $X_i(t)$ from adding standard normal random errors to two mean spectra extracted from our real application data. Each mean spectrum was used for generating 30 covariate functions, and so the sample size was 60. The true coefficient function $\beta(t)$ is plotted as the black solid line in Figure~\ref{fig:sflr_sim2b}, which also resembles the coefficient function estimate from our application (Figure~\ref{fig:sflr_real1}). The intercept $\alpha$ was chosen such that the binary groups generated from model \eqref{eq:flmodel} had roughly equal sizes. 
The tuning parameters $\lambda$ and $\gamma$ were selected respectively from the grids $(4, 6, 8, \ldots, 20)$ and $(2,3,4)*10^{-6}$. The SFLR estimate of $\beta$, as plotted as the red solid line in Figure~\ref{fig:sflr_sim2b}, clearly did a good job in distinguishing null regions from non-null-regions while providing a reasonably good estimate at the non-null-regions. 

\section{Real Data}
\label{se:real}
Hemodialysis is the most common treatments for patients with end stage renal disease. In a hemodialysis treatment session, typically about 4 hours long, fresh dialysate is continually circulated through the dialyzer to remove metabolic waste products from patients' blood. In our hemodialysis study, waste dialysate samples (containing metabolic wastes) were collected at 10 min, 60 min, 120 min, 180 min, and 240 min (the end) of the session. Each sample was divided into 10 portions and each portion was analyzed by a Raman spectrometer (Peakseeker Pro 785, Agiltron Inc., Woburn, MA) to produce a raw spectrum. Therefore, a total of 50 raw Raman spectra with 10 spectra associated with each time point were generated for the session.

\begin{figure}[]
    \centering
    \includegraphics[scale=0.75]{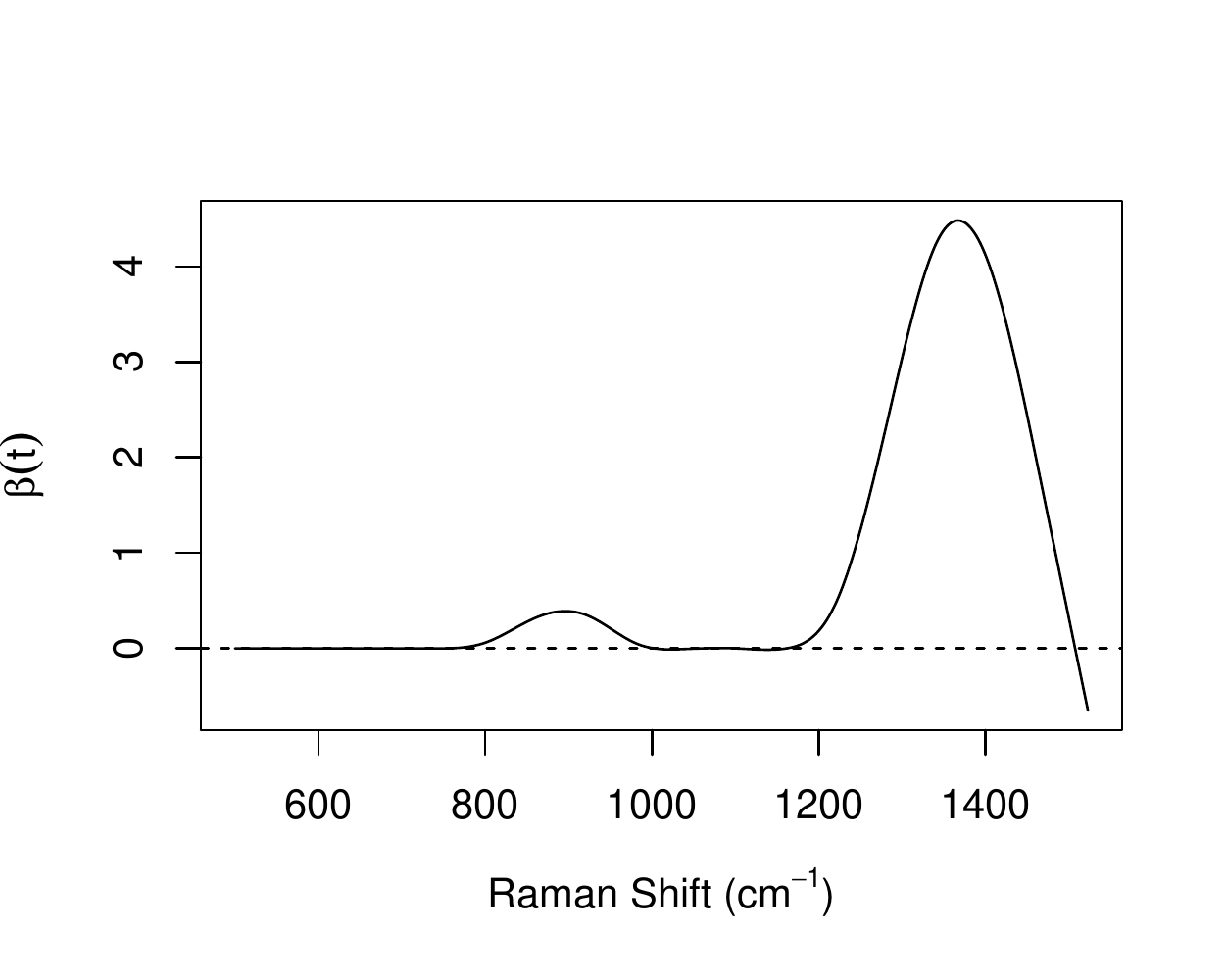}
    \caption{SFLR estimate of $\beta(t)$ from hemodialysis spectra.}
    \label{fig:sflr_real1}
\end{figure}

The two-sample test from \citet{Horvath2013} was applied to those spectra and found significant difference in two groups of spectra as shown in the top panel of Figure \ref{fig:ch4real}. Their respective mean spectra are plotted in the bottom panel of Figure \ref{fig:ch4real}, where we can see that the main difference lie in the regions around 900 cm$^{-1}$ and 1400 cm$^{-1}$. We applied the proposed SFLR method to the two groups of spectra with the parameters $\lambda$ and $\gamma$ selected by the BIC respectively from the grids $(1,2,3,\ldots, 10)$ and $(0.55,1)*10^{-5}$. The estimated $\hat{\beta}(t)$ is plotted in Figure \ref{fig:sflr_real1}. $\hat{\beta}(t)$ is mostly 0 except for the two regions $[774\text{cm}^{-1}, 996\text{cm}^{-1}]$ and $[1168\text{cm}^{-1}, 1523\text{cm}^{-1}]$. It suggests that chemicals whose Raman peaks fall within these two regions have the most significant contribution to differentiating the two groups of waste dialysate spectra. This finding is consistent with the mean spectral plot in Figure~\ref{fig:ch4real}. 

\section{Conclusion}
\label{se:con}
Despite a rich literature on generalized functional linear models, none has considered the case with a locally sparse coefficient function that is practically meaningful. Motivated by a biomedical study on hemodialysis monitoring, we propose a locally sparse functional logistic regression method by applying an $L_1$-norm local sparsity penalty and a roughness penalty to the coefficient function. The problems boils down to the optimization of a doubly-penalized likelihood where local sparsity and smoothness are enforced through their respective penalties. A Newton-Raphson procedure is proposed for computation. Our simulation assessment and application of the proposed SFLR method to hemodialysis spectra have shown its capability of identifying null region(s) of the coefficient function and generating a smooth estimate of the function on the non-null region(s).

Our method only considers functional data with a binary response. It can be easily modified to the more general case where the response comes from an exponential family of distributions. For example, a generalization of the work in \citet{du2014} can be obtained through the addition of a local sparsity penalty to their penalized likelihood that involves only a roughness penalty.

In this paper we have demonstrated the proposed locally sparse logistic regression method through its application to hemodialysis monitoring. It has much wider applications in many scientific or medical studies using Raman spectroscopy. For example, we are also planning on applying it to characterization of urine from patients with end-stage kidney disease \citep{eskd} and screen of bladder cancer \citep{bladder}. Besides early detection of cancers and monitoring of medical procedures or treatment effects, other potential medical applications include classification and differentiation of diseased tissues from normal ones and determination of molecular compositions of diseased tissues or pathogens.


\bibliographystyle{imsart-nameyear}
\bibliography{ref.bib}          
%
%

\end{document}